\documentclass{revtex4}

 \usepackage{graphicx, color}

\usepackage{amssymb,amsmath}

\begin{document}


\title{Finite dipolar hexagonal columns on piled layers of triangular lattice}

\author{Katsuyoshi Matsushita}
\email{kmatsu@issp.u-tokyo.ac.jp}
\affiliation{Institute for Solid State Physics, University of Tokyo, 5-1-5 Kashiwanoha, Kashiwa-shi, Chiba 277-8581, Japan. }
\author{Ryoko Sugano}
\affiliation{Advanced Research Laboratory, Hitachi, Ltd.,
   1-280 Higashi-Koigakubo, Kokubunji-shi, Tokyo 185-8601.}
\author{Akiyoshi Kuroda}
\affiliation{Institute for Solid State Physics, University of Tokyo, 5-1-5 Kashiwanoha, Kashiwa-shi, Chiba 277-8581, Japan. }
\author{Yusuke Tomita}
\affiliation{Institute for Solid State Physics, University of Tokyo, 5-1-5 Kashiwanoha, Kashiwa-shi, Chiba 277-8581, Japan. }
\author{Hajime Takayama}
\affiliation{Institute for Solid State Physics, University of Tokyo, 5-1-5 Kashiwanoha, Kashiwa-shi, Chiba 277-8581, Japan. }
\received{}
\revised{}
\accepted{}

\begin{abstract}
 We have investigated, by the Monte Carlo simulation, spin systems which represent moments of arrayed magnetic nanoparticles interacting with each other only by the dipole-dipole interaction. In the present paper we aim the understanding of finite size effects on the magnetic nanoparticles arrayed in hexagonal columns cut out from the close-packing structures or from those with uniaxial compression. In columns with the genuine close-packing structures, we observe a single vortex state which is also observed previously in finite 2-dimensional systems. On the other hand in the system with the inter-layer distance set $1/\sqrt{2}$ times of the close-packing one, we found ground states which depend on the number of layers. The dependence is induced by a finite size effect and is related to a orientation transition in the corresponding bulk system.
\end{abstract}


\maketitle

In recent years, systems consisting of arrayed single-domain ferromagnetic nanoparticles receive much attention of many researchers because of its possibility as a high storage density\cite{Chou}. 
In such array systems the dipole-dipole interaction plays an important role on their magnetic properties. 
In particular the dipole-dipole interaction originates strong finite size effects.
For example the peculiar ``from-edge-to-interior freezing'' of a finite dipolar cube is found in our recent work\cite{1}. 
In the present paper, we aim to clarify finite size effects due to the dipole-dipole interaction on magnetic nanoparticles arrayed in hexagonal columns cut out from piled triangular lattice with the close-packing structures (FCC and HCP) or with those uniaxially-compressed in a direction of the 3-fold axis.

We perform simulated annealing of the dipolar columns by the heat-bath Monte Carlo method. 
We employ the well-known Hamiltonian of a dipole-dipole interacting system,
\begin{center}
\begin{eqnarray}
H = \frac{D}{2a^3}\sum_{<i,j>} {\vec S}_i \cdot \frac{1-{\vec e}_{ij} \otimes {\vec e}_{ij} }{r_{ij}^3} \cdot {\vec S}_j, \label{Ham}
\end{eqnarray}
\end{center}
where $\vec S_i$'s denote classic Heisenberg spins which represent magnetic moments of nanoparticles. 
$D$ is a coupling constant and $a$ a lattice constant.
Other effects in nanoparticle systems are omitted because we focus only on those by the dipole-dipole interaction.
The number of piled layers in the columns examined, $N_{\rm L}$, is from 1 to 15. Each layer contains about 100 sites.

\begin{figure}[t]
  \begin{center}
   \resizebox{0.5\textwidth}{!}{\includegraphics{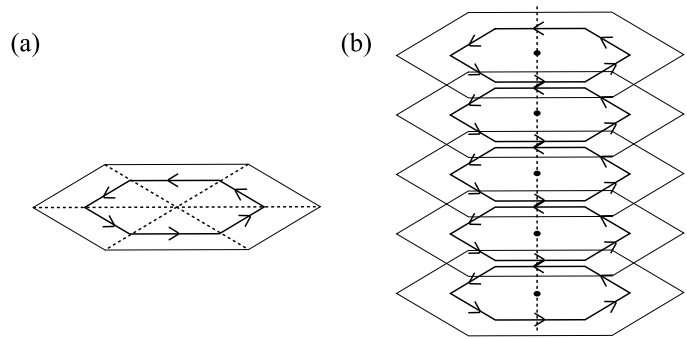}}
  \end{center}
  \caption{The schematic image of the single vortex state in the single-layer (a) and  multi-layer (b) systems. The solid and dashed lines denote the directions of the spins and the domain boundary between ferromagnetic domains, respectively.}
\label{fig1}
\end{figure}

As well known\cite{Belobrov,Vedmedenko}, 
spin configurations consist of in-plane vortices at low temperatures in finite dipolar systems on the triangular lattice.
In the present simulation, spins in the ground state of a single-layer system also form an in-plane single vortex shown in Fig.~\ref{fig1}(a).
We can explain the origin of the vortex state as below.
In the bulk system the ground state is in-plane ferromagnetic one\cite{Rosenbaum}. 
In the state, Hamiltonian (\ref{Ham}) possesses the global symmetry of $O(2)$, which corresponds to the global in-plane spin-rotation. 
Thus the direction of the ferromagnetic order is not fixed in the bulk system.
In a finite system, on the other hand, the symmetry around an edge is reduced into $Z_2$ by the anisotropy of the dipole-dipole interaction.
On the edge the ferromagnetic order is in the direction parallel to the edge.
A fixed ferromagnetic domain extends from each edge because of this symmetry reduction.
In fact, the from-edge-to-interior freezing is observed in the single-layer system. 
The ferromagnetic domains continuously merge with neighboring domains and form the single vortex state as a whole.
This symmetry reduction is similar to the one found in the dipolar cube\cite{1}.
We can also expect it around domain walls observed in large systems\cite{Vedmedenko}.

\begin{figure}[t]
  \begin{center}
   \resizebox{0.3\textwidth}{!}{\includegraphics{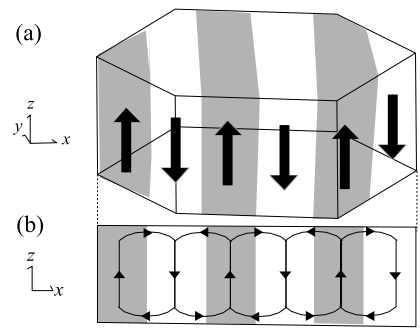}}
  \end{center}
  \caption{The schematic image of the multi-ferromagnetic-domain structure (a) and of the convective flow pattern of ferromagnetic spin chains (b). The arrows denote the direction of spins, respectively.}
\label{fig3}
\end{figure}

The nature of the single layer is kept in multi-layer columns with the genuine close-packing structures. 
In fact a single vortex state is also observed as a whole as shown in Fig.~\ref{fig1}(b). 
We next examine columns with the inter-layer distance set $1/\sqrt{2}$ times of the close-packing one, i.e., columns on close-packing lattices compressed in the direction of their 3-fold axis. 
In the system of 2 layers, the single vortex state is also observed. 
However we have found a multi-ferromagnetic-domain structure, which is shown in Fig.~\ref{fig3}(a), when $N_{\rm L}$ is more than 2. 
In its ferromagnetic domain, 
spins align in the direction perpendicular to layers and in the direction anti-parallel to those of neighboring domains except for those around top and bottom layers. 
On the top and bottom layers, 
spins align so as to connect smoothly between those in neighboring domains. Spin configuration as a whole form a convective flow pattern of ferromagnetic spin chains between the top and bottom layers as shown in Fig.~\ref{fig3}(b). 
In the multi-layer hexagonal column, the number of the domain decreases from 6 to 3 as $N_{\rm L}$ increases from 3 to 15. 

To interpret the results observed numerically,
let us first consider the corresponding bulk system. 
Its ground state is considered to consist of multi-ferromagnetic domain\cite{Kittel}.
We focus attention on a domain in the ground state.
Now we assume that the domain is enough large that contributions of boundary energy and of inter-domain interaction are negligibly small as compared to that of the intra-domain interaction energy, $E_{\rm D}$. 
Because of the symmetry of Hamiltonian (\ref{Ham}) $E_{\rm D}$ is described by magnetization of the domain,
\begin{eqnarray}
 E_{\rm D} = - J (M_x^2 + M_y^2) - \Delta M_z^2,
\end{eqnarray}
where $M_x$, $M_y$ and $M_z$ denote the magnetization of the domain with $\sum_{\alpha} M_{\alpha}^2= M^2$, $J$ and $\Delta$ anisotropy constants.
The $z$-direction is set in the direction of the 3-fold axis.
This formula holds in all lattices examined in the present paper, i.e., FCC, HCP and those with the compression in the $z$-direction.
In these systems if $J > \Delta$, the in-plane $O(2)$-symmetric state with $M_z = 0$ appears. If $J < \Delta$, the out-of-plane state with $M_z = \pm M$ appears.
The bulk system is, thereby, expected to exhibit a first order transition when the ratio $J/\Delta$ is change.
As known previously,  $J = \Delta$ is for genuine FCC\cite{LT}. We assume from our observation described above that $J \sim \Delta$ is for genuine HCP and that $J < \Delta$ for those compressed $1/\sqrt{2}$ times at least.

Next let us introduce upper and lower boundaries into the bulk single domain. If $N_{\rm L}$ is 2, i.e., the system consists of only two boundary triangular lattice layers, 
the situation of $J > \Delta$ is expected to hold due to the strong anisotropy on the boundaries and the in-plane $O(2)$-symmetric state is preferred.
When we next introduce the lateral boundary with the hexagonal shape, the in-plane single vortex state appears as in the single layer described above.
As the number of layers increases, the effective fields
that spins inside a column feel come closer to those of the bulk single
domain. In columns consisting of multi-layer on the genuine
close-packing lattice, the situation of $J > \Delta$ is kept, and the
single vortex state as a whole is realized (Fig.~\ref{fig1}(b)), because $J \sim \Delta$ at most in the bulk single domain.
In multi-layer systems with the compression, on the other hand, 
the situation of $J < \Delta$ is expected as we have assumed above, and 
the out-of-plane state is preferred. The consequence is the convective
flow pattern state shown in Fig.~\ref{fig3}, which associates with zero
total magnetization. The latter result is a very common property of
dipolar systems, because of which we have to consider a multi-domain
state even in bulk (infinite) systems. 

To conclude, the finite size effect, combined with the peculiar anisotropic nature of the dipole-dipole interaction induces the orientation change in
the ground state of dipolar columns consisting of triangular lattice
layers as the interlayer distance or the number of layers is changed.

\end{document}